\documentstyle[preprint,aps,epsf,floats,fleqn]{revtex}
\tighten
\def\be{\begin{equation}}
\def\ee{\end{equation}}
\def\nn{\nonumber}
\def\mdot{\!\cdot\!}
\def\li2{\mbox{Li}_2}
\begin{document}

\begin{titlepage}

  \preprint{UJ--TPJU 18/96} 
  \title{Perturbative QCD corrections to inclusive lepton
    distributions from semileptonic $b \rightarrow c\tau\bar\nu_\tau$
    decays \thanks{Talk given by L.M. at the XXXVI Cracow School
    of Theoretical Physics, Zakopane, June 1996. Work supported in
    part by KBN grant 2P30207607.}}
  \author{Marek Je\.{z}abek$^{a,b}$ and Leszek Motyka$^c$}
  \address{\medskip
    (a) Institute of Nuclear Physics, Kawiory 26a,PL-30055 Cracow,
        Poland \\
    (b) Institute of Physics, Silesian University, Katowice,
        Poland \\
    (c) Institute of Physics, Jagellonian University, Reymonta 4,
          PL-30-059 Cracow, Poland. } \vspace{2em}
  \date{September 1996}
  \maketitle
\begin{abstract}

  Perturbative corrections of the order of $\alpha_s$ to inclusive
  double differential lepton distribution from $b$~quark decay are
  considered.
  A perturbative correction to the charged lepton energy spectrum
  has been calculated for an
  arbitrary charged lepton mass.
  The perturbative contribution suppresses the partial
  rate but almost does not change the shape of energy
  distribution.
  Applications of our result to semileptonic B meson decays
  are briefly discussed.

\end{abstract}
\thispagestyle{empty}
\end{titlepage}

\section{Introduction}

The precise determination of weak mixing angles $V_{cb}$ and $V_{ub}$
is a demanding task. In spite of progress in this field they still
remain ones of the worse known parameters of the Standard Model. The
uncertainties which appear here are both of an experimental and
theoretical origin. The relatively large theoretical errors mainly
reflect the lack of quantitative knowledge about the structure of
hadrons and QCD higher order perturbative corrections to the
amplitudes of weak decays of $b$~quarks. The most valuable source of
information about the weak mixing angles are the semileptonic decays
of $B$ and $\bar{B}$ mesons. The leptons in the final state do not
interact strongly and the process is less affected by unknown QCD
effects than a hadronic decay. Furthermore pseudoscalar $B$~mesons are
the simplest bottom hadrons.  The $b$~quark mass is about 5~GeV and
thus it exceeds roughly ten times typical energy scales which
characterize the infrared dynamics in the hadrons. Moreover the
presence of this mass justifies the perturbative treatment of most of
the processes involving the $b$~quark. The simple facts have given
rise to a quantitative description of dynamics of hadrons containing
heavy quarks (Heavy Quark Effective Theory
\cite{VS,PW,IsgW,EH,Grin,Geor}). In the framework of HQET many
observables describing heavy hadrons may be expressed as a power
series in $1/m_b$.  In particular it was shown in Ref.~\cite{CGG},
that the inclusive lepton distributions from a bottom hadron decay may
be treated in such a way.  It follows from the operator product
expansion (OPE) that a matrix elements which should be evaluated to
derive the distributions may be expanded into a series of local
operators characterizing the decaying bound state. The very advantage
of this approach is that subsequent unknown non-perturbative matrix
elements are suppressed by increasing powers of $m_b$. The leading
term corresponds to a parton contribution to the process. As argued in
Ref.~\cite{CGG} the next-to-leading term vanishes. The $1/m_b ^2$
corrections have been calculated by for a case of massless
\cite{IB,MWB} and \cite{nonp1,nonp2,nonp3} massive lepton. Recently
also the third order terms have become known \cite{GK}.

The first order perturbative QCD corrections to the inclusive lepton
distributions in a process of decay: $b \rightarrow ql\bar\nu$ are as
important as the HQET corrections for the corresponding $B$ decay.
They have been evaluated \cite{JK,CJ} for the vanishing lepton
mass.  In the case of a non-zero lepton mass only a differential
distribution of the lepton pair invariant mass is known to the first
order in strong coupling constant \cite{CJK}.
In the present article we present our recent calculation
of the first order QCD correction to the double
differential inclusive lepton distribution from $b$~decay with a
massive lepton in the final state. The complete analytical
result and details of the calculation will be published elsewhere
\cite{JM}.
Here we give results for
the perturbative correction to the $\tau$ lepton
energy spectrum which has been
obtained by numerical integration of this double differential
distribution.

\section{Kinematical variables}

The purpose of this section is to define the kinematical variables which are
used in this paper. We describe also the constraints imposed on these variables
for three and four-body decays of the heavy quark.

The calculation is performed in the rest frame of the decaying $b$ quark. Since
the first order perturbative QCD corrections to the inclusive process are taken
into account, the final state can consist either of produced quark $c$, lepton
$\tau$ and $\tau$ anti-neutrino or of the three particles and a real gluon. The
four-momenta of the particles are denoted in the following way: $Q$ for
$b$~quark, $q$ for $c$~quark, $\tau$ for the charged lepton, $\nu$ for the
corresponding anti-neutrino and $G$ for the real gluon. By the assumption that
all the particles are on-shell, the squares of their four-momenta are equal to
the squares of masses:
\be
Q^2 = m_b ^2, \qquad q^2 = m_c ^2 , \qquad \tau^2 = m_\tau ^2 , \qquad
\nu^2 = G^2 = 0.
\ee
The four-vectors $P = q + G$ and $W = \tau + \nu$ characterize
the quark--gluon system and the virtual $W$ boson respectively. We define a set
of variables scaled in the units of mass of heavy quark $m_b$:
\be
\varrho = {m_c ^2 \over m_b ^2} , \qquad \eta = {m_\tau ^2 \over m_b ^2},\qquad
x = {2 E_\tau \over m_b ^2}, \qquad t = {W^2 \over m_b ^2}, \qquad
z = {P^2 \over m_b ^2}.
\ee
We introduce light-cone variables describing the charged lepton:
\be
\tau_\pm = {1\over 2} \left( x \pm \sqrt{x^2 - 4\eta} \right)
\ee
The system of $c$ quark and real gluon is characterized by the following
quantities:
\begin{eqnarray}
P_0 (z)   &=& {1 \over 2}(1-t+z),   \nn \\
P_3 (z)   &=& \sqrt{P_0 ^2 - z} = {1\over 2} [1+t^2+z^2-2(t+z+tz)]^{1/2},\nn \\
P_\pm (z) &=& P_0 (z) \pm P_3 (z),  \nn \\
{\cal Y}_p (z) &=& {1 \over 2} \ln {P_+ (z) \over P_- (z) } =
                 \ln { P_+ (z) \over \sqrt{z} }  \nn \\
\end{eqnarray}
where $P_0 (z)$ and $P_3 (z)$ are the energy and length of the momentum vector
of the system in $b$ quark rest frame, ${\cal Y}_p (z)$ is the corresponding
rapidity.
Similarly for virtual $W$:
\begin{eqnarray}
W_0 (z) &=& {1 \over 2}(1+t-z),  \nn \\
W_3 (z) &=& \sqrt{W_0 ^2 - t} = {1\over 2} [1+t^2+z^2-2(t+z+tz)]^{1/2},  \nn \\
W_\pm (z) &=& W_0 (z) \pm W_3 (z),   \nn \\
{\cal Y}_w (z) &=& {1 \over 2} \ln {W_+ (z) \over W_- (z) } =
                 \ln { W_+ (z) \over \sqrt{t} },   \nn \\
\end{eqnarray}
>From kinematical point of view the three body decay is a special case of the
four body one with vanishing gluon four-momentum, what is equivalent to
$z=\varrho$. It is convenient to use in this case the following variables:
\[
p_0 = P_0 (\varrho) = {1\over 2} (1 - t + \varrho ), \qquad
p_3 = P_3 (\varrho) = \sqrt{p_0 ^2 - \varrho},
\]
\[
p_\pm = P_\pm (\varrho) = p_0 \pm p_3,                 \qquad
w_\pm = W_\pm (\varrho) = 1 - p_\mp .
\]
\be
Y_p  = {\cal Y}_p (\varrho ) = {1 \over 2} \ln {p_+ \over p_-}, \qquad
Y_w  = {\cal Y}_w (\varrho ) = {1 \over 2} \ln {w_+ \over w_-}.
\ee
We express also the scalar products which appear in the calculation by the
variables $x$, $t$ and $z$:
\be
\begin{array}{ll}
Q \mdot P \,= {1\over 2}   (1+z-t) & \tau\mdot\nu \, = {1\over 2}(t - \eta)   \\
Q \mdot \nu \, = {1\over 2} (1-z-x+t)\hspace{3em} &
                                     \tau\mdot P \, = {1\over 2}(x-t-\eta)    \\
Q \mdot \tau \, = {1\over 2} x     & \nu\mdot\tau \, = {1\over 2}(1-x-z+\eta) \\
\end{array}
\ee
All of the written above products are scaled in the units of the mass of
$b$~quark.

The allowed ranges of $x$ and $t$ for the three-body decay
are given by following inequalities:
\be
2\sqrt{\eta} \leq x \leq 1 + \eta - \varrho = x_{max}, \qquad
\label{xbound}
\ee
\be
t_1 = \tau_- \left( 1 -  {\varrho \over 1 - \tau_-} \right)
\leq t \leq
\tau_+ \left( 1 -  {\varrho \over 1 - \tau_+} \right) = t_2
\ee
(a region A).
In the case of the four-body process the available region of the phase space
is larger than the region~A. The additional, specific for the four body decay area
of the phase space is denoted as a region~B. Its boundaries are given by the
formulae:
\be
2\sqrt{\eta} \leq x \leq x_{max}, \qquad
\eta \leq t \leq t_1
\label{xyps}
\ee
We remark, that if the charged lepton mass tends to zero than
the region B vanishes.

One can also parameterize the kinematical boundaries of $x$ as
functions of~$t$. In this case we obtain for the region A:
\be
\eta  \leq  t  \leq  (1-\sqrt{\varrho})^2, \qquad
w_- + {\eta \over w_-} \leq x \leq w_+ + {\eta \over w_+},
\label{yxps}
\ee
and for the region B:
\be
\eta  \leq  t  \leq  \sqrt{\eta} \left( 1 - {\varrho\over 1 - \sqrt{\eta}}
                                                  \right), \qquad
2\sqrt{\eta}  \leq  x  \leq  w_- + {\eta \over w_-}.
\ee

The upper limit of the mass squared of the $c$-quark --- gluon system is
in both regions given by
\be
z_{max} = (1-\tau_+)(1-t/\tau_+),
\ee
whereas the lower limit depends on a region:
\be
z_{min} = \left\{
\begin{array}{ll}
\varrho                  & \mbox{\rm in the region A} \\
(1-\tau_-)(1-t/\tau_-)   & \mbox{\rm in the region B.} \\
\end{array} \right.
\ee

\section{Evaluation of the QCD corrections}

The QCD corrected differential rate for
$b \rightarrow c + \tau^- + \bar{\nu}$
reads:
\be
d\Gamma = d\Gamma_0 + d\Gamma_{1,3} + d\Gamma_{1,4},
\ee
where
\be
d\Gamma_0 = G_F ^2 m_b ^5 |V_{CKM}| ^2 {\cal M}_{0,3} ^-
           d{\cal R}_3 (Q;q,\tau,\nu) / \pi^5
\ee
in Born approximation,
\be
d\Gamma_{1,3} = {2 \over 3}\alpha_s G_F ^2 m_b ^5 |V_{CKM}|^2 {\cal M}_{1,3} ^-
           d{\cal R}_3 (Q;q,\tau,\nu) / \pi^6
\ee
comes from the virtual gluon contribution and
\be
d\Gamma_{1,4} = {2 \over 3}\alpha_s G_F ^2 m_b ^5 |V_{CKM}|^2 {\cal M}_{1,4} ^-
           d{\cal R}_4 (Q;q,\tau,\nu) / \pi^7
\ee
describes a real gluon emission. $V_{CKM}$ is the Cabbibo--Kobayashi--Maskawa
matrix element associated the $b$ to $c$ or $u$ quark weak transition.
Lorentz invariant $n$-body phase space is defined as
\be
d{\cal R}_n(P;p_1, \ldots , p_n ) =
\delta^{(4)} (P - \sum p_i) \prod_i { d^3 {\bf p}_i \over 2 E_i}
\ee
In Born approximation the rate for the decay into three body final state is
proportional to the expression
\be
{\cal M}_{0,3} ^- = F_0 (x,t) =
4 q \mdot \tau \; Q \mdot \nu \;= (1 - \varrho - x + t )(x - t - \eta),
\ee
where the quantities describing the $W$~boson propagator are neglected.
Interference between virtual gluon exchange and Born amplitude yields:
\begin{eqnarray}
{\cal M}_{1,3} ^- & = &
        - [
   \; q \mdot\tau\; Q \mdot \nu  \; H_0 + \varrho\; Q\mdot\nu\; Q\mdot\tau\; H_+
 + \; q \mdot \nu\; q \mdot \tau \; H_- +  \nn\\
  & &    + {1 \over 2} \varrho \; \nu \mdot \tau \; ( H_+ + H_-)
         + {1 \over 2} \eta \varrho  \; Q \mdot \nu \; ( H_+ - H_- + H_L )
         - {1 \over 2} \eta \; q \mdot \nu \; H_L ] , \nn  \\
\end{eqnarray}
where
\begin{eqnarray}
H_0 & = & 4(1-Y_p p_0/p_3 ) \ln \lambda_G + (2p_0/p_3)
          \left[ \li2 \left( 1 - {p_- w_- \over p_+ w_+ } \right) \right. \nn\\
    &   & - \left. \li2 \left( 1 - {w_- \over w_+} \right) - Y_p (Y_p+1) +
          2(\ln\sqrt\varrho + Y_p)(Y_w + Y_p) \right] \nn\\
    &   & + [2p_3 Y_p + (1 - \varrho - 2t) \ln \sqrt \varrho ] / t + 4, \nn\\
H_\pm & = & {1 \over 2} [ 1 \pm (1-\varrho) / t ] Y_p / p_3 \pm
                    {1 \over t} \ln\sqrt\varrho , \nn \\
H_L & = &  {1 \over t} ( 1 -\ln\sqrt\varrho) + {1- \varrho \over t^2}
           \ln\sqrt\varrho + {2 \over t^2} Y_p p_3 + {\varrho\over t}
            {Y_p \over p_3}. \nn\\
\end{eqnarray}
In ${\cal M}_{1,3} ^-$
infrared divergences are regularized by
a small mass of gluon denoted by $\lambda_G$.
According to Kinoshita--Lee--Naunberg theorem, the infrared divergent part
should cancel with the infrared contribution of the four-body decay amplitude
integrated over suitable part of the phase space.

The rate from real gluon emission is proportional to
\be
{\cal M}^- _{1,4} =
                    {{\cal B}^- _1 \over (Q\mdot G)^2 } -
                    {{\cal B}^- _2 \over Q\mdot G \; P\mdot G} +
                    {{\cal B}^- _3 \over (P\mdot G)^2 } ,
\ee
where
\begin{eqnarray}
 {\cal B}_1 ^- & = & \,
  q \mdot \tau \; [\, Q \mdot \nu\; (Q \mdot G \, - 1) + \, G \mdot \nu \,  -
            \, Q \mdot \nu\; Q \mdot G \, +\, G \mdot \nu\; Q \mdot G \, ],\nn\\
 {\cal B}_2 ^- & = & \,
  q \mdot \tau \; [\, G \mdot \nu \; Q \mdot q \, - \, q\mdot\nu \; Q\mdot G\, +
  \, Q \mdot \nu \; (\, q \mdot G \, - \, Q \mdot G\, - 2\, q \mdot Q\,)]+ \nn\\
               &   & + \, Q \mdot \nu \; (\, Q \mdot \tau \; q \mdot G \,
                     - \, G \mdot \tau\;  q \mdot Q\, ), \nn\\
 {\cal B}_3 ^- & = &   Q \mdot \nu \; (\, G \mdot \tau \; q \mdot G \, -
                       \varrho \; \tau \mdot P \, ). \nn\\
\end{eqnarray}

Integrating and adding all the contributions one arrives at the
following double differential distribution of leptons:
\be
{d\Gamma \over dx\, dt} =
\left\{
\begin{array}{ll}
12 \Gamma_0 \left[ F_0 (x,t) - {2\alpha_s \over 3\pi } F_1 ^A (x,t) \right] &
\mbox{for $(x,t)$ in A}, \\
12 \Gamma_0 {2\alpha_s \over 3\pi} F_1 ^B (x,t) & \mbox{for $(x,t)$ in B} \\
\end{array}
\right.
\label{main}
\ee
where
\be
\Gamma_{0} =  {G_F ^2 m_b ^5 \over 192\pi^3} |V_{CKM}|^2,
\ee
\be
F_0 (x,t) = (1 - \varrho - x + t )(x - t - \eta),
\ee
and the functions $F_1 ^A (x,t)$, $F_1 ^B (x,t)$ describe the
perturbative correction in the regions $A$ and $B$.
Explicite formulae for $F_1 ^A (x,t)$ and $F_1 ^B (x,t)$
will be given in \cite{JM}.
The factor of 12 in the formula (\ref{main}) is introduced to meet widely used
\cite{nonp1,CJK,MV} convention for $F_0 (x)$ and $\Gamma_0$.

The obtained results were tested by comparison with
earlier calculations.
One of the cross checks was arranged by fixing the
mass of the produced lepton to zero.
Our results are in this limit algebraically identical
with those for the massless charged lepton \cite{JK,CJ}.
On the other hand one can numerically integrate the calculated
double differential distribution over $x$, with the limits given by the
kinematical boundaries:
\be
\int_{2\sqrt{\eta}} ^{w_+ + \eta / w_+ } {d\Gamma\over dx\, dt}
(x,t;\varrho,\eta) = {d\Gamma \over dt} (t;\varrho,\eta)
\label{testint}
\ee
Obtained in such a way differential
distribution of $t$ agrees with recently published \cite{CJK}
analytical formula describing this distribution. This test is particularly
stringent because one requires two functions of three variables
($t,\varrho$ and $\eta$) to be numerically equal for any values of the arguments.
We remark, that for higher values of $t$
only the region A contributes to the integral (\ref{testint}) and for lower
values of $t$ both regions~A and B contribute.
This feature of the test is very helpful ---
the formulae for $F_1 (x,t)$, which are different for the
regions A and B can be checked separately.

\begin{figure}
\hbox{
\epsfxsize = 200pt
\epsfysize = 200pt
\epsfbox[36 366 453 765]{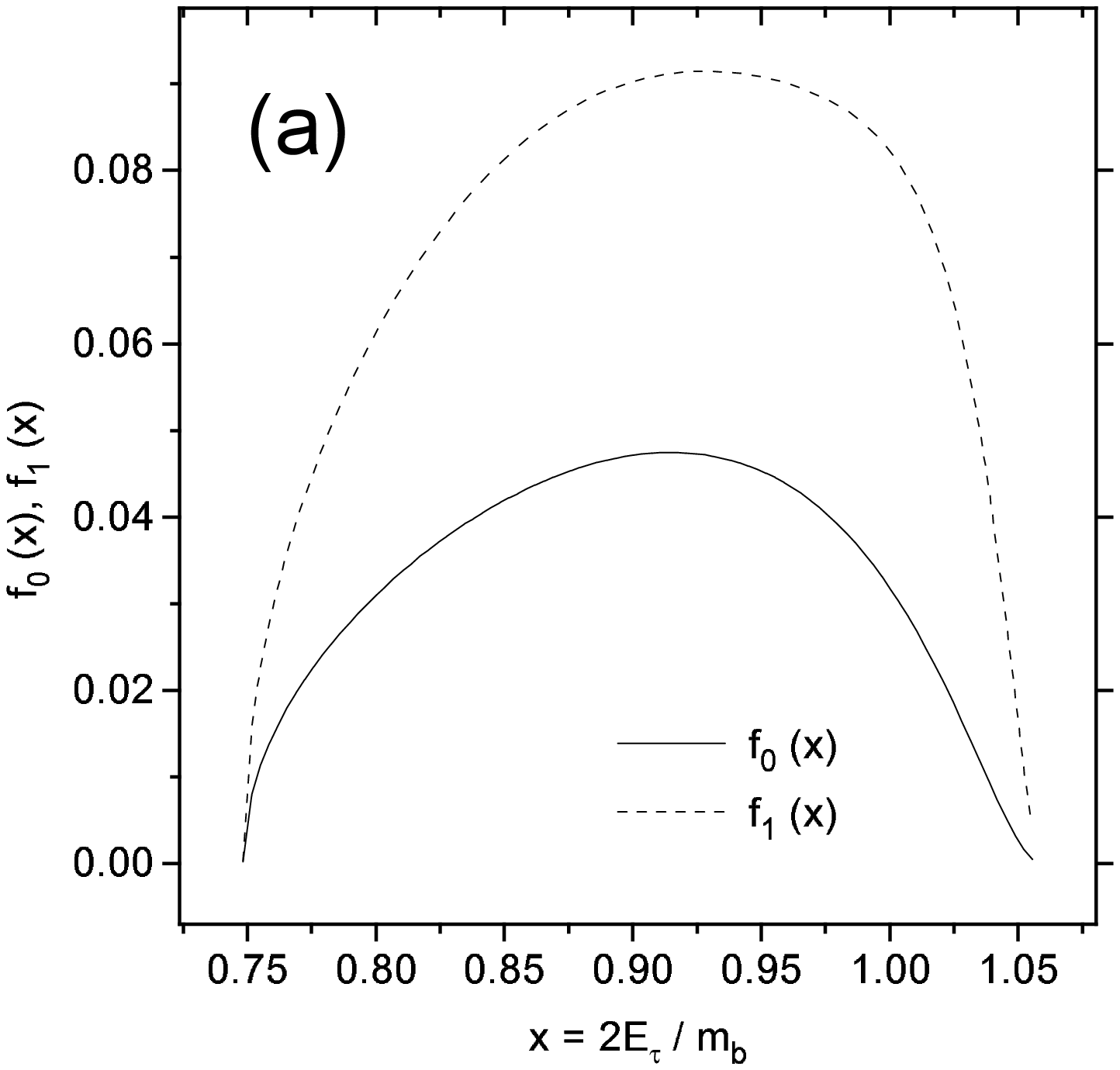}
\epsfxsize = 200pt
\epsfysize = 200pt
\epsfbox[36 366 453 765]{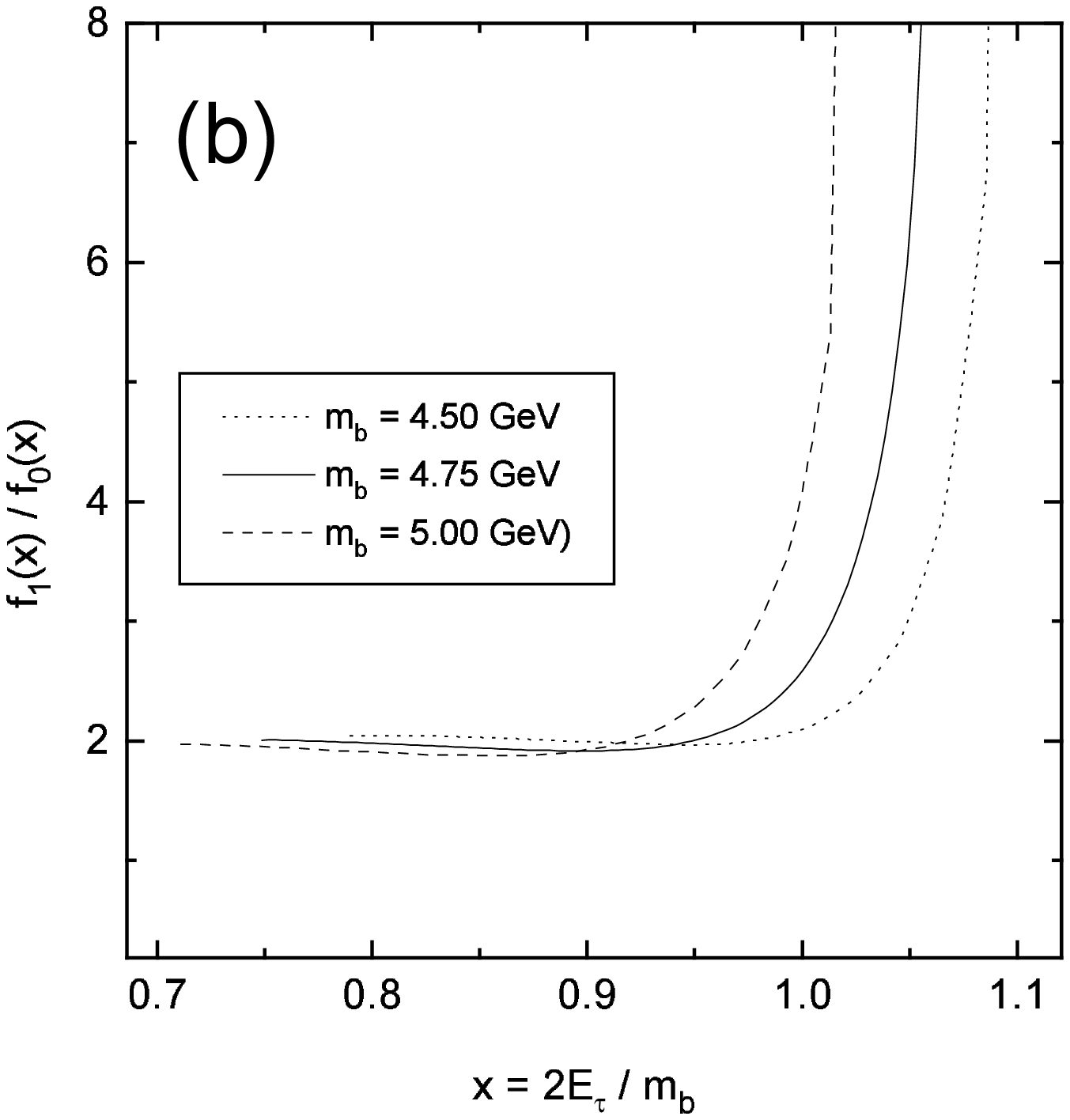} \vspace{1em} }
\caption{(a) The distributions $f_0 (x)$, $f_1 (x)$  and (b) the
ratio $f_1 (x) / f_0 (x)$ for the pole mass of the $b$ quark $m_b =
4.5$~GeV (dotted), $m_b = 4.75$~GeV (solid) and $m_b = 5.0$~GeV (dashed).}
\end{figure}

\section{Differential distribution of $\tau$ energy}

The point of interest to check how the QCD corrections change
energy spectrum of the charged lepton. This aim may be reached by integration
of the double differential lepton distribution over the lepton pair invariant
mass:
\be
{d\Gamma \over dx} = \int_{\eta} ^{t_2} {d\Gamma \over dx\,dt} \, dt,
\ee
where $t_2$ is the upper kinematical boundary for $t$ given the formula
(\ref{yxps}). The decomposition of the resulting distribution into the Born
term and the perturbative QCD correction yields in a natural way definitions of
functions $f_0 (x)$ and $f_1 (x)$:
\be
{d\Gamma \over dx} =
12 \Gamma_0 \left[ f_0 (x) - {2\alpha_s \over 3\pi } f_1 (x) \right].
\ee
The analytical formula for $f_0 (x)$ reads
\begin{eqnarray}
f_0 (x) &=& 2\sqrt{x^2 - 4\eta}
\left\{ x_0 ^3 [x^2 - 3x(1 + \eta) + 8\eta] + \right.  \nn \\
 & & \hspace{100pt}\left. + x_0 ^2 [-3x^2 + 6x(1+\eta) - 12\eta ] \right\} \nn ,\\
\end{eqnarray}
where following \cite{nonp1} we introduced
\be
x_0 = 1 - \varrho /(1+\eta-x) .
\ee
An equivalent expression for $f_0 (x)$ is
\begin{eqnarray}
f_0 (x) &=& {1\over 6} x \sqrt{ x^2 - 4\eta }
         \left( \frac{ x_{max} - x }{1+\eta -x}  \right) ^2
    [ 3 (1+\eta ) - 2x + \varrho - 4\eta / x +                 \nn \\
 & & \hspace{105pt} + 2\varrho (1+\eta -4\eta / x) / (1+\eta -x)],  \nn \\
\end{eqnarray}
where $x_{max}$ is given by (\ref{xbound}). The latter formula
clearly exhibits the behavior of $f_0 (x)$ for $x$ close to
the upper kinematical limit.

The integration of $F^{A,B} _1 (x,t)$ was performed numerically
for different masses of $b$-quark with fixed $m_b - m_c = 3.4$~GeV and
$m_\tau = 1.777$~GeV. The  functions  $f_0 (x)$ and
$f_1 (x)$ for $m_b = 4.75$~GeV are plotted on Fig.~1a and the ratios
$f_1 (x) / f_0 (x)$ for three different realistic values of
$m_b$ are plotted on Fig.~1b. As can be easily seen the ratios have logarithmic
singularities at the upper end of the spectra. Such a behavior would lead to a
inconsistence. The standard solution to problems of this kind is an
exponentiation which yields well known Sudakov form factor \cite{FJMW}.
Far from the end point the ratio of the correction term to the leading one
is almost constant and close to~2. It means that the perturbative correction
changes rather the normalization than the shape of lepton energy distribution.

\begin{figure}[hbpt]
\epsfxsize = 350pt
\epsfysize = 290pt
\hspace{2cm}
\epsfbox[85 366 517 775]{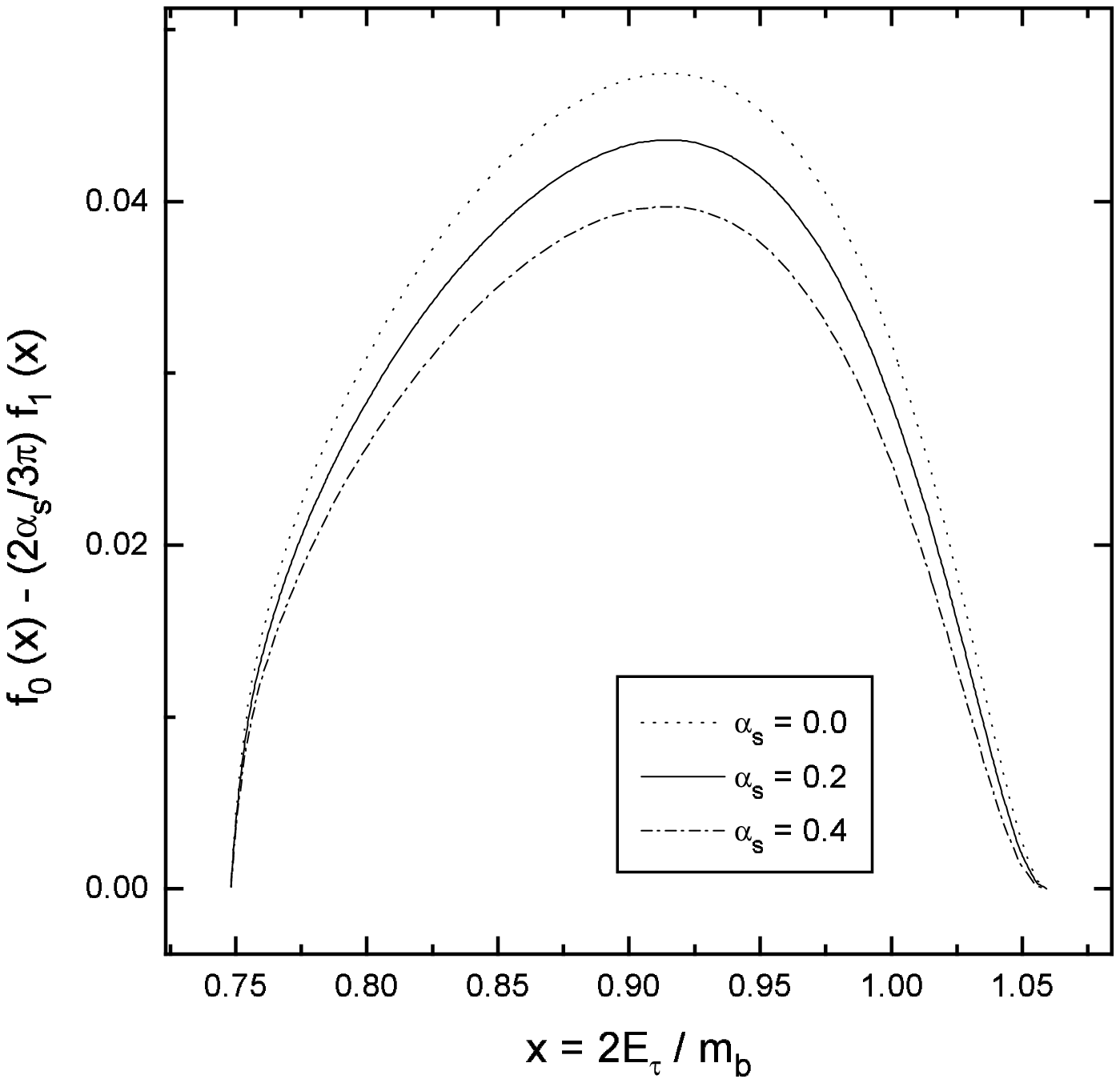}
\caption{The QCD corrected $\tau$ lepton spectrum from the $b$ quark
decay for different values of $\alpha_s$. The mass of $b$ quark is chosen as
4.75~GeV.}
\end{figure}
\
The obtained distributions of the scaled charged lepton energy for
$m_b=4.75$~GeV with and without perturbative QCD corrections are
shown in
Fig.~2. The strong coupling constant was chosen as $0.2$ and $0.4$ since the
energy scale for this process in not known until the second order QCD
corrections are evaluated. The value of $\alpha_s$ for this decay is expected
to lay between the two numbers.

The knowledge of perturbative corrections to lepton energy is essential for
fixing HQET parameters, especially $\lambda_1$ and $\bar\Lambda$
\cite{IB,MWB}. Especially analysis of moments of the lepton energy spectrum
and other quantities involving integration over the energy distribution
appeared particularly valuable for this purpose \cite{GKLW} as was earlier
suggested in Refs.~\cite{CJK,MV}.

\section{Conclusions}

The first order QCD corrections to the double differential inclusive lepton
distributions from $b$~quark semileptonic decay have been
calculated for a massive fermion in the final state.
Non-trivial cross checks of the final the result have been performed.
We remark that including a real gluon radiation on the parton level yields a
increase of the phase space available in the decay process.
The QCD corrected $\tau$ energy spectrum has been obtained. The effect of the
correction may be estimated as about 10\% of the magnitude of uncorrected
distributions.

The presented above results can be utilized to improve an analysis of
semileptonic decays of beauty hadrons with a $\tau$ in the final state.
Thus the values of involved weak mixing angles may be fixed more exactly.
The decrease of theoretical uncertainty increases the sensitivity to hypothetic
deviations from the Standard Model \cite{Kal,GL,GHN} which should have to be
particularly distinct in the case of the heaviest family.  The better
understanding of the perturbative QCD effects allows one to perform more
stringent tests of  HQET predictions \cite{nonp1,nonp2,nonp3} and narrow the
error bars for HQET parameters. Moreover one can extract more precisely some
information about masses of quarks and strong coupling constant from the data.
Finally, the process that we considered may appear a background for other
processes so precise theoretical knowledge about the process is valuable.
At present however the statistics of measured
$b \rightarrow c(u)\tau\bar\nu_\tau$
transitions  is rather low and ten-percent effects are not seen. Probably the
application of provided here formulae to the expected data from $B$-factories
will be really fruitful.

\end{document}